\documentclass{PoS}
\usepackage{amsmath,bbm}

\newenvironment{nscenter}
 {\parskip=0pt\par\nopagebreak\centering}
 {\par\noindent\ignorespacesafterend}

\let\OLDthebibliography\thebibliography
\renewcommand\thebibliography[1]{
  \OLDthebibliography{#1}
  \setlength{\parskip}{0pt}
  \setlength{\itemsep}{0pt plus 0.0ex}
}

\newcommand{\nobracket}{}
\newcommand{\nocomma}{}

\title{Lattice Hamiltonian approach to the Schwinger model: further results from the strong coupling expansion}

\ShortTitle{Lattice Hamiltonian approach to the Schwinger model -- SCE}

\author{\speaker{Marcin Szyniszewski}\thanks{M.S. is fully funded by EPSRC, NoWNano DTC grant number EP/G03737X/1.}\\
        Physics Department, Lancaster University, Lancaster, LA1 4YB, UK\\
				NoWNano DTC, University of Manchester, Manchester, M13 9PL, UK\\
        E-mail: \email{mszynisz@gmail.com}}

\author{Krzysztof Cichy\thanks{K.C. acknowledges very useful discussions with Mari Carmen Ba\~{n}uls, Karl Jansen and Hana Saito.}\\
        J. von Neumann Institute for Computing (NIC), DESY, Platanenallee 6, 15738 Zeuthen, Germany\\
				Faculty of Physics, Adam Mickiewicz University, Umultowska 85, 61-614 Pozna\'{n}, Poland\\
        E-mail: \email{krzysztof.cichy@desy.de}}

\author{Agnieszka Kujawa-Cichy\\
        Institut f\"ur Theoretische Physik, Goethe-Universit\"at, 60438 Frankfurt/Main, Germany\\
				E-mail: \email{cichy@th.physik.uni-frankfurt.de}}

\abstract{
  We employ exact diagonalization with strong coupling expansion to the
  massless and massive Schwinger model. New results are presented for the 
	ground state energy and scalar mass gap in the massless model, which improve
	the precision to nearly $10^{-9} \%$. We also investigate the chiral condensate
  and compare our calculations to previous results available in the
  literature. Oscillations of the chiral condensate which are present while
  increasing the expansion order are also studied and are shown to be directly
  linked to the presence of flux loops in the system.}

\FullConference{The 32nd International Symposium on Lattice Field Theory,\\
		23-28 June, 2014\\
		Columbia University New York, NY}

\begin{document}
\section{Introduction}

The Schwinger model {\cite{Schwinger1962}}, i.e. QED in 1+1 dimensions, has become a toy model for gauge
theories.
It is simple enough to analytically extract its behavior in various limits
{\cite{Crewther1980}}, but it still exhibits such interesting phenomena as
chiral symmetry breaking and quark confinement {\cite{Hamer1982}}.

The model has been studied thoroughly using various methods
{\cite{Crewther1980}--\cite{Banuls2014pos}}. Recently, we were able to show that the Hamiltonian approach
still can give results that are comparable or sometimes exceed the precision of other methods
{\cite{Cichy2013}}. In this article, we continue that study by showing updated 
results and expand it to another observable -- the chiral order parameter (chiral condensate).

The outline of this paper is as follows. In Section 2 we shortly recall the
Schwinger model and Section 3 describes the method we use -- exact
diagonalization with strong coupling expansion. Our results, a comparison to previous findings and
a thorough description of chiral condensate oscillations
are included in Section 4, which is then followed by Section 5 -- the summary.

\section{The Schwinger model}

The Hamiltonian of the massive Schwinger model on lattice, in the
Kogut-Susskind discretization {\cite{Kogut1975,Banks1976}} and after the
Jordan-Wigner transformation {\cite{Jordan1928}} to the spin space is :
\begin{equation}
  \hat{\mathcal{H}} = \frac{1}{2 a} \sum_{n = 1}^M \left( \sigma^+ (n) e^{i
  \theta (n)} \sigma^- (n + 1) + \text{h.c.} \right) + \frac{m}{2} \sum_{n =
  1}^M (1 + (- 1)^n \sigma^3 (n)) + \frac{g^2 a}{2} \sum_{n = 1}^M L^2 (n),
  \label{schwinger}
\end{equation}
where $a$ -- lattice spacing, $m$ -- fermion mass, $g$ -- coupling, $M$ -- system size.
$\sigma^i(n)$ are Pauli matrices
operating in spin space, $L (n)$ is related to electric field $L (n) = E
(n) / g$ and operates in ladder space $L | l \rangle = l | l \rangle, l =
0, \pm 1, \pm 2, \ldots$, and $e^{\pm i \theta (n)}$ are rising and lowering
operators in this ladder space.

We will be interested in assessing the chiral condensate $\Sigma$, which is
given by:
\begin{equation}
  \frac{\Sigma}{g} = \frac{\sqrt{x}}{2 M} \langle 0 | \sum_{n = 1}^M (- 1)^n \sigma^3
  (n) | 0 \rangle,
	\label{condensate}
\end{equation}
where $| 0 \rangle$ is the ground state of the Hamiltonian and $x = \frac{1}{a^2 g^2}$. The
theoretical prediction for the massless
model is $\nobracket \nobracket \nobracket \Sigma/g |_{m = 0} = -
e^{\gamma}/{2 \pi^{3 / 2}} \approx - 0.1599288$.

\section{Method}

The Hamiltonian (\ref{schwinger}) can be rewritten in the dimensionless form
as:
\begin{equation}
  \hat{W} = \frac{2}{a g^2} \hat{\mathcal{H}} = \hat{W}_0 + x \hat{V}
\end{equation}
with $\hat{W}_0$ including the mass part and the electric field part and
$\hat{V}$ including the hopping. If $x$ is small (strong coupling), then we
can treat $\hat{W}_0$ as an unperturbed Hamiltonian and $\hat{V}$ as a
perturbation.

We employ the exact diagonalization (ED) approach with the strong coupling
expansion (SCE) introduced by Hamer {\cite{Hamer1979}} and initially used for
the Schwinger model in Refs. {\cite{Crewther1980,Hamer1982}}. SCE truncates the Hamiltonian, by selecting
only those states that are
connected with the ground state up to a specific order of perturbation. Then,
the truncated Hamiltonian is finite and thus amenable to ED.

In Ref. {\cite{Cichy2013}} we suggested to use a very high order $N$ of SCE, for which the eigenvalues
are saturated, i.e. do not change when further increasing $N$, up to machine precision. The
saturation was present for observables such as ground state energy and scalar/vector mass gaps for the
massless model. The same method will be used to determine another observable -- the chiral condensate.

\section{Results and comparison}
\subsection{Ground state energy and scalar mass gap}

\begin{figure}[t!]
  \centering
  \resizebox{12.5cm}{!}{\includegraphics[angle=270]{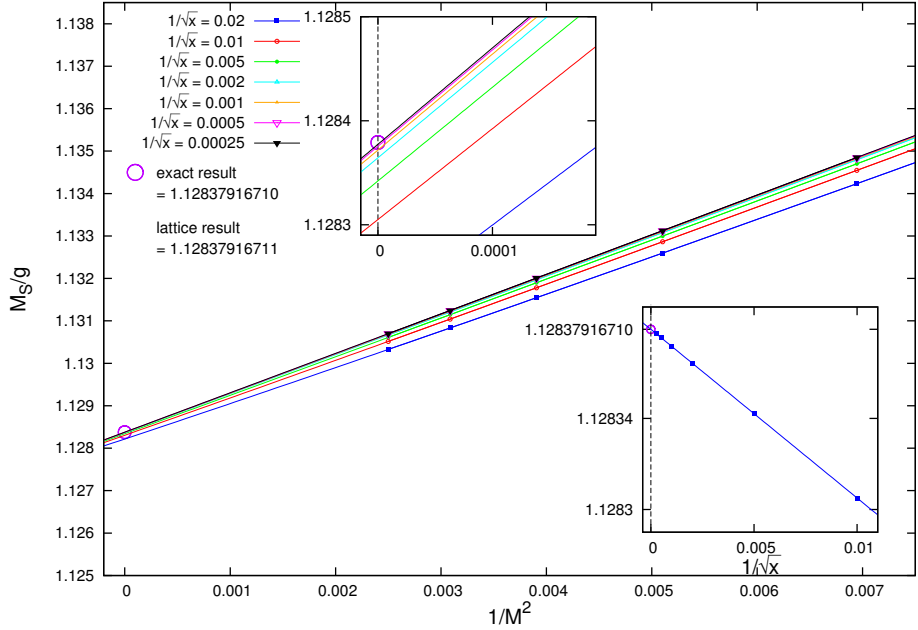}}
  \caption{
	  Infinite volume extrapolation ($M\rightarrow\infty$) of the scalar mass gap. Top inset
		shows a zoomed in view. Bottom inset shows the continuum extrapolation ($1/\sqrt{x}\to
		0$).\label{massgapplot}
		}
\end{figure}

First, we present updated results for the ground state energy $E_0$ and scalar mass
gap $M_S/g$ for the massless Schwinger model, obtained by increasing the maximal system size (with
respect to Ref.~\cite{Cichy2013}) and thus being able to get closer to the continuum limit (increase
maximal $x$). We show the continuum limit extrapolation of the 
scalar mass gap in Fig.~\ref{massgapplot}, while Tab.~\ref{energytable} 
presents the comparison to our previous result, which shows significant 
improvement.

\begin{table}[t!]
  \small
	\begin{nscenter}
		\begin{tabular}{lllll}
			Observable & $E_0$ & (error) & $M_S / g$ & (error)\\
			\hline
			Previous calculation {\cite{Cichy2013}} & $- 0.3183098827$ & $1.1 \cdot
			10^{- 6} \%$ & $- 1.1283791668$ & $2.9 \cdot 10^{- 8} \%$\\
			This work & $- 0.3183098860$ & $6.5 \cdot 10^{- 8} \%$ & $- 1.12837916711$
			& $1.3 \cdot 10^{- 9} \%$\\
			Exact value & $- 0.3183098862$ & $-$ & $- 1.12837916710$ & $-$
		\end{tabular}
  \end{nscenter}
	\caption{Comparison with previous results \cite{Cichy2013} for ground state energy $E_0$
	  and scalar mass gap $M_S/g$.\label{energytable}}
\end{table}

\subsection{Chiral condensate}
Our results for the chiral condensate in the massless case are summarized in
Fig.~\ref{chiralmassless}.
The inset shows an example of our infinite volume extrapolation, using an exponential ansatz normally
expected for periodic boundary conditions away from the critical point ($x\rightarrow\infty$).
The results in infinite volume are then shown in the main plot and used to extrapolate to the continuum
limit. It can be shown even in the free theory that the approach to the continuum limit is linear in the
lattice spacing with logarithmic corrections (and higher-order corrections). In the plot, we show such
fit and compare it to the purely linear one. Indeed, the correct result in the continuum limit is
obtained only if the logarithmic corrections are taken into account.

{

\begin{figure}[t!]
  \centering
  \resizebox{8cm}{!}{\includegraphics[angle=270]{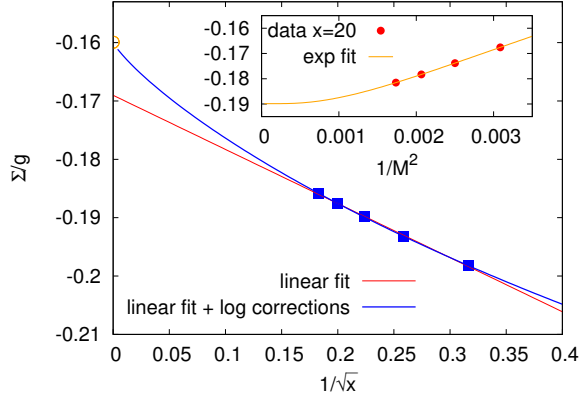}}
  \caption{Chiral condensate for the massless case -- continuum limit
  estimation. Inset shows an example of the infinite volume limit estimation 
	for $x = 20$ with exponential fit.\label{chiralmassless}}
\end{figure}
}

\begin{table}[t!]
	\small
	\begin{nscenter}
		\begin{tabular}{lllll}
			$x$ & $m/g$ & SCE+ED & MPS {\cite{Banuls2013pos}} & Difference\\
			\hline
			20 & 0 & $- 0.189879$ & $- 0.190253$ & 0.000374\\
			25 & 0 & $- 0.187519$ & $- 0.187969$ & 0.000450\\
			30 & 0 & $- 0.185830$ & $- 0.186208$ & 0.000378\\
			cont. & 0 & $- 0.1600 (17)$ & $- 0.159930 (8)$ & \\
			cont. & 0.125 & $- 0.0906(23)$ & $- 0.092023 (4)$ & 
		\end{tabular}
	\end{nscenter}
  \caption{Comparison of SCE+ED and MPS results {\cite{Banuls2013pos}} for the chiral condensate in the
massless and massive ($m/g = 0.125$) model. For the massive
  model, a logarithmic divergence had to be subtracted.}
\label{comparechiralmassless}
\end{table}

The comparison with Matrix Product States (MPS) results {\cite{Banuls2013pos}} is
shown in Tab.~\ref{comparechiralmassless}. Interestingly, the infinite volume
limit values for SCE+ED are always a bit smaller than those for MPS. This might suggest that our
exponential fitting function is subject to power-law corrections that dominate the behavior close
to the continuum limit (cf. infinite volume scaling for the scalar mass gap -- Fig.~\ref{massgapplot}).
The continuum limit result is in agreement with MPS, but it has much lower accuracy, due to the
fact that MPS allows to study much larger system sizes and much larger values of $x$.


The massive case was also investigated, where a logarithmic divergence has to be subtracted
(this divergence is present already in the free case). The fermion mass tends to
increase finite volume effects.
Once again, though our findings are consistent with previous work, they have 
much less precision.

\subsection{Oscillations of the chiral condensate}

\subsubsection{Description of the oscillations}\label{description}

When one starts increasing the SCE order $N$, for some values of $x$ a problem appears with estimating
the saturated values. Examples are shown in Fig.~\ref{oscillations}. We can clearly see that for small
values of $x$ (away from the continuum limit), oscillations are very small, almost non-existent.
However, when we approach the continuum limit and thus $x$ is very big, the
oscillations are clearly visible and we cannot directly extract the saturated
value.
{
\setlength{\floatsep}{5pt plus 0pt minus 0pt} 
\setlength{\textfloatsep}{7pt plus 0pt minus 0pt} 
\setlength{\intextsep}{7pt plus 5pt minus 5pt} 

\begin{figure}[t!]
  \centering
  \resizebox{15cm}{3.5cm}{\includegraphics{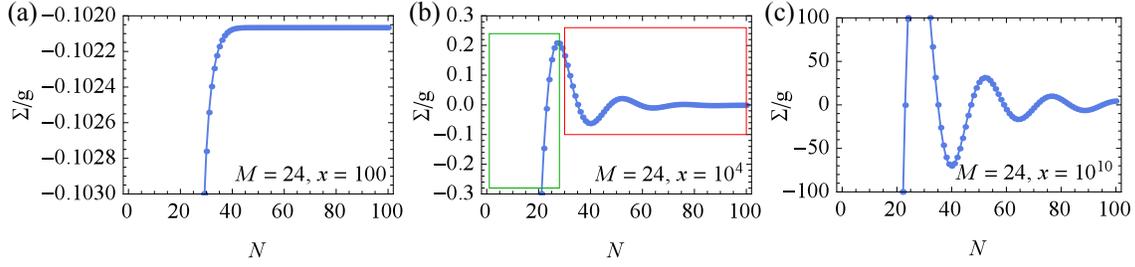}}
  \caption{Chiral condensate when increasing the SCE order $N$ for different
  couplings: (a) $x = 10^2$, (b) $x = 10^4$, (c) $x =
  10^{10}$.\label{oscillations} On plot (b) we can see a tail, marked in green
  and oscillations, marked in red.}
\end{figure}
}
Interestingly, the plots show two distinct regions: the tail at $N \lesssim M$ and the oscillations
which have a very specific period. For now, we will ignore the tail part and try to
describe the oscillations by fitting them to a chosen function. Our ansatz
includes an oscillating part (sine), a modulation (decreasing function of
$N$) and a constant shift (saturated value of $\Sigma$). It was found that the
modulation part can be best described as a function of $\frac{1}{N^3}$ for
large $x$ and as an inverse exponential function of $N$ for small $x$. Thus,
the final fitting function is:
\begin{equation}
  \Sigma (N) /g = \Sigma (N \rightarrow \infty)/g + a \left( \frac{b}{N^3} + e^{-
  \alpha N} \right) \sin \left( \frac{2 \pi}{T} N + \varphi \right),
\end{equation}
where $\Sigma (N \rightarrow \infty)/g, a, b, \alpha, T$ and $\varphi$ are the
fitting parameters. The data was always fitted starting from the second
extremum of $\Sigma (N)/g$, so that the effects of the tail part are greatly
diminished.
\begin{figure}[t!]
  \centering
	  \resizebox{!}{5.6cm}{\includegraphics{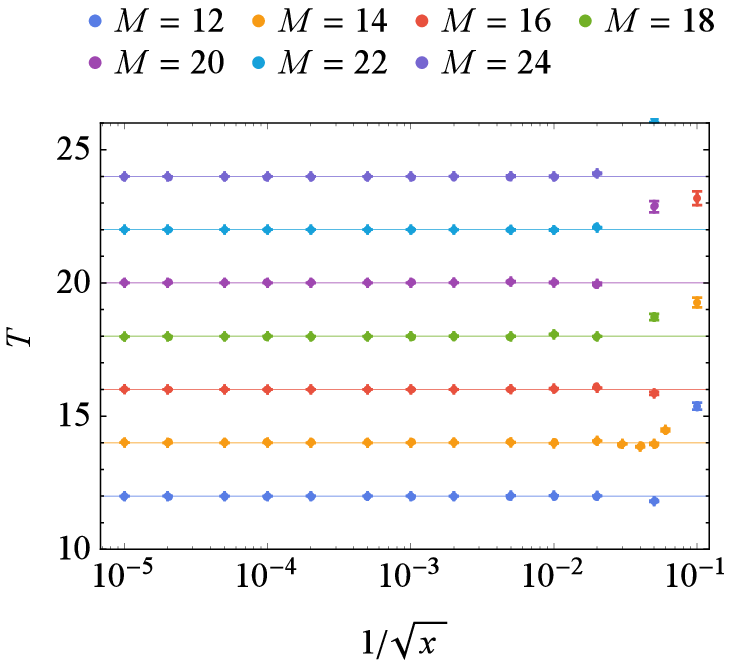}}
	  \resizebox{!}{6.0cm}{\includegraphics{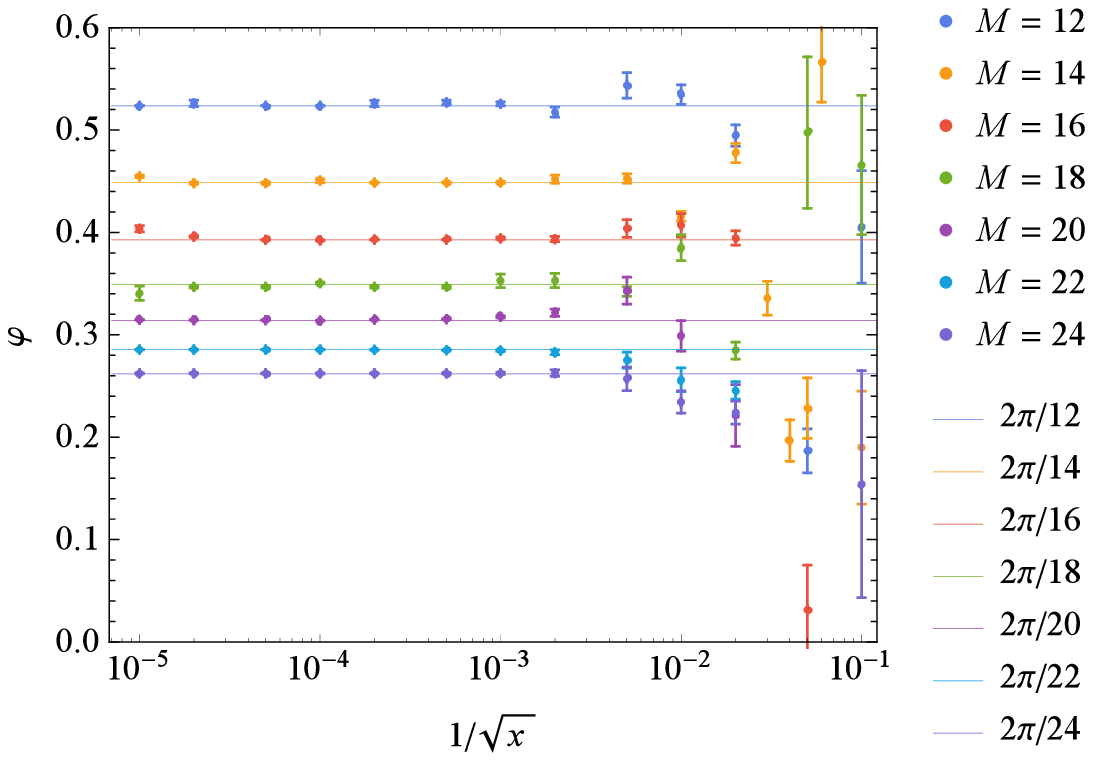}}
  \caption{Period (left) and phase (right) of oscillations.\label{Tphi} The
  lines on the $T$ plot indicate $T = 12, 14, \ldots, 24$.}
\end{figure}

The period and the phase of oscillations are shown in Fig.~\ref{Tphi}. We can
clearly see that close to the continuum limit (large $x$), we have the
following dependencies:
\begin{equation}
  T = M \nocomma, \hspace{1em} \varphi = \frac{2 \pi}{M}.
\end{equation}
However, these equations seem to be invalid for small $x$. This is due to the fact that for small $x$ the
oscillations are too small to be well-described by our ansatz. On the phase $\varphi$ plot, we can see
that errors grow for small $x$, which also indicates this problem.


\begin{figure}[t!]
	\begin{minipage}[b]{6.45cm}
		\resizebox{6.45cm}{!}{\includegraphics{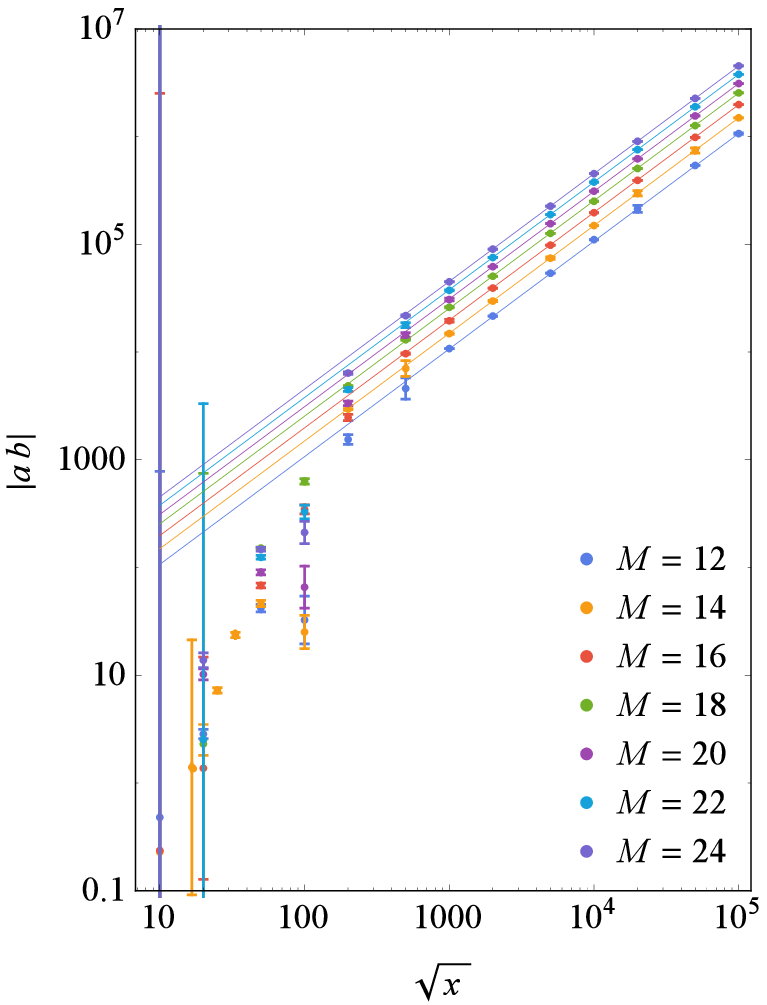}}
		\caption{Coefficient in front of $\frac{1}{N^3}$. The solid lines are the
			linear fits to $y = a x$ function, which on the log-log will always be 
			parallel to each other.\label{ab}}
	\end{minipage}\hspace{0.4cm}
	\begin{minipage}[b]{8.2cm}
		\resizebox{8.1cm}{!}{\includegraphics{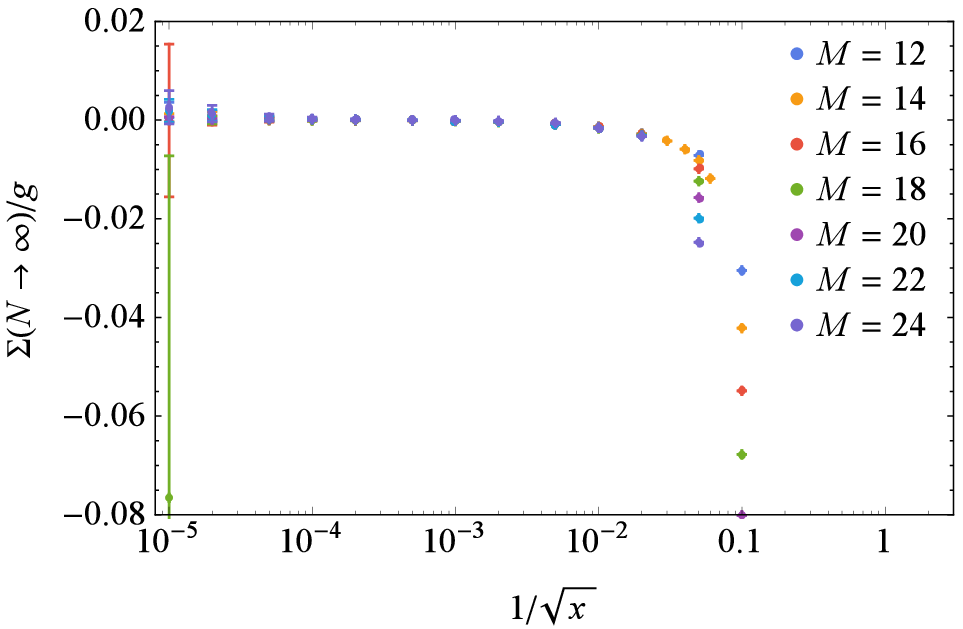}}
		\caption{Saturated values of the chiral condensate, $\Sigma (N \rightarrow
			\infty)/g$.\label{satur} Notice that the naive extrapolation to the continuum
			limit would yield $\Sigma/g \rightarrow 0$. \protect\linebreak}
	\end{minipage}
\end{figure}

The coefficient in front of $\frac{1}{N^3}$ was also investigated and it is shown
in Fig.~\ref{ab}. We are taking the absolute value of the coefficient, due
to the sine function being antisymmetric ($\sin x = - \sin (- x)$). The data seems
to indicate that this coefficient is proportional to $\sqrt{x}$:
\begin{equation}
  | a b | = A (M) \cdot \sqrt{x}.
\end{equation}
However, on the log-log plot we can clearly see that for small $x$, this
dependence is again invalid.


Fig.~\ref{satur} shows saturated values of the chiral condensate obtained
from fitting of the oscillations. Interestingly, for a very large $x (\sim
10^{10})$, there are very big errors present -- this is due to huge amplitude
of oscillations.

It would seem that the continuum limit value $\Sigma (M \rightarrow \infty, x
\rightarrow \infty)/g$ is zero. However, this invalid result is obtained due to
large finite volume effects: we firstly have to take the infinite volume limit
which is quite problematic with our approach, if we work too close to the continuum limit. Only after
taking the infinite volume limit, we can take the continuum limit by
fitting the infinite volume limit data and extracting the $x \rightarrow \infty$
value.

\subsubsection{Final fitting ansatz}

Using relationships extracted in section \ref{description}, we can rewrite the
final fitting ansatz as:
\begin{equation}
  \Sigma (N, M, x)/g = \Sigma (N \rightarrow \infty, M, x)/g + \left( A (M)
  \frac{\sqrt{x}}{N^3} + B (M, x) e^{- \alpha (M, x) N} \right) \sin \frac{2
  \pi}{M} (N + 1).
\end{equation}
We can immediately see that for $N = k M - 1$ the sine function will reach the
middle point of oscillations. So, if there is any residual function that we
omitted in our fitting ansatz, it will be present at those points. Therefore,
we suggest to use values for $N = k M - 1$ to extract the saturated value of
the chiral condensate by fitting those specific plot points.

\subsubsection{Connection to the number of flux loops in the system}

Every time $N = k M, k \in \mathbbm{Z}$, states generated in the SCE procedure
will include the next positive $(L (n) = + 1, + 2, + 3, \ldots)$ and the next
negative $(L (n) = - 1, - 2, - 3, \ldots)$ flux loop in the system. For
example, for $N = 2 M$, the Hamiltonian will include flux loops $L (n) = - 2,
- 1, + 1, + 2$. Thus, we can see that the value for SCE order $N$ must be very
close to the value for SCE order $N + M$, because both systems have very
similar structure except for the number of included flux loops. We therefore
conclude that the period of the oscillations being a constant $M$ is a manifestation of
the flux loops in the system.

Now, we can also see that the physical interpretation of the tails in Fig.~\ref{oscillations} is the
absence of the flux loops in the Hamiltonian, which is true for $N < M$.

\section{Summary and outlook}

We presented updated results for the ground state energy and scalar mass gap for
the massless Schwinger model. The precision of scalar mass gap calculation was
increased up to almost $10^{- 9} \%$, which, to our knowledge, is the most precise lattice
result ever obtained. It is, however, of minor practical importance, because it can not be systematically
extended to other observables or non-zero fermion mass. As such, it should be considered an interesting
peculiarity of the spectrum of the massless case.

We have also shown findings for the chiral condensate for both massless and massive
model. Our results indicate large finite volume effects that make it impossible to achieve lattice
spacings as small as when using the MPS method. Note, however, that the lattice spacings that can be used
are still much smaller than when using standard Monte Carlo methods (where typically $\beta\equiv
x\leq10$).

Finally, we have described our findings concerning chiral condensate oscillations when increasing the SCE
order and we have shown that
they can be directly linked to the number of flux loops in the system.

\end{document}